\documentclass[preprint,aps,12pt,showpacs,nofootinbib,tightenlines]{revtex4}
\usepackage{mathrsfs}
\usepackage{amsmath}
\usepackage{amssymb}
\usepackage{epsfig}
\usepackage{graphicx}
\newcommand{\psl}{ P \hspace{-2.4truemm}/ }
\newcommand{\nsl}{ n \hspace{-2.2truemm}/ }
\newcommand{\vsl}{ v \hspace{-2.2truemm}/ }
\newcommand{\esl}{ \epsilon \hspace{-1.5truemm}/ }

\def\be{\begin{eqnarray}}
\def\en{\end{eqnarray}}
\def\non{\nonumber\\}

\def\prd{{Phys. Rev. D}~}
\def\prl{{ Phys. Rev. Lett.}~}
\def\plb{{ Phys. Lett. B}~}
\def\npb{{ Nucl. Phys. B}~}

\newcommand{\acp}{{\cal A}_{CP}}
\begin{document}
\title{Branching Ratio and Polarization of $B \to a_1(1260)(b_1(1235))\rho(\omega, \phi)$ Decays in the PQCD Approach }
\author{Zhi-Qing Zhang
\footnote{Electronic address: zhangzhiqing@haut.edu.cn} } 
\affiliation{\it \small  Department of Physics, Henan University of
Technology, Zhengzhou, Henan 450052, P.R.China } 
\date{\today}
\begin{abstract}
Within the framework of perturbative QCD approach, we study the charmless two-body decays into final states involving one axial-vector (A), $a_1(1260)$ or $b_1(1235)$,
and one vector (V), namely $\rho(\omega,\phi)$. Using the decays constants and the light-cone distribution amplitudes for these mesons derived from the QCD sum rule
method, we find the following results: (a) Except the decays $\bar B^0\to  a^{0}_1\rho^0(\omega)$, other tree-dominated
decays $B\to a_1\rho(\omega)$ have larger branching ratios, at the order of $10^{-5}$. (b)Except the decays $\bar B\to b^+_1\rho^-$ and $B^-\to b^0_1\rho^-$, other
$B\to b_1\rho(\omega)$ decays have smaller branching ratios, at the order of $10^{-6}$. (c) The decays $B\to a_1(b_1) \phi$ are highly suppressed and have very small
branching ratios, at the order of $10^{-9}$. (d) For the decays $\bar B^0 \to a_1^0\rho^0$ and $B^-\to b_1^-\rho^0$, their two transverse polarizations are larger than
their longitudinal polarizations, which are about $43.3\%$ and $44.9\%$, respectively. (d) The two transverse polarizations have near values in the decays $B\to a_1\rho(\omega)$,
while have large differences in some of $B\to b_1\rho(\omega)$ decays. (e) For the decays $B^-\to  a^{0}_1\rho^-, b^{0}_1\rho^-$ and
$\bar B^0\to  b^{0}_1\rho^0, b^{0}_1\omega$, where the transverse polarization fractions range from $4.7$ to $7.5\%$, we calculate their direct
CP-violating asymmetries with neglecting the transverse polarizations and find that those for two charged decays have smaller values, which are about $11.8\%$ and $-3.7\%$,
respectively.
\end{abstract}

\pacs{13.25.Hw, 12.38.Bx, 14.40.Nd}
\vspace{1cm}

\maketitle


\section{Introduction}\label{intro}
In general, the mesons are classified in $J^{PC}$ multiplets. There
are two types of orbitally excited axial-vector mesons, namely
$1^{++}$ and $1^{+-}$. The former includes $a_1(1260), f_1(1285),
f_1(1420)$ and $K_{1A}$, which compose the $^3P_1$-nonet, and the
latter includes $b_1(1235), h_{1}(1170), h_1(1380)$ and $K_{1B}$, which
compose the $^1P_1$-nonet. There is an important character for these
axial-vector mesons except $a_1(1260)$ and $b_1(1235)$, that is each different flavor state can mix
with one another, which comes from the other nonet meson or the same nonet one.
There is not mix between $a_1(1260)$ and $b_1(1235)$ because of
the opposite C-parities. They do not also mix with others. So compared with other axial-vector mesons, these two mesons should have less uncertainties about their inner structures.

Like decay modes $B\to VV$, the charmless decays $B\to AV$ also have three polarization states and so are expected to have rich physics.
In many $B\to VV$ decays, the informations
on branching ratios and polarization fractions among various helicity amplitudes have been studied by many authors \cite{yli,ali,hwhuang,beneke}.
Through polarization studies, some underling helicity structures of the decay mechanism are proclaimed.
They find that the polarization fractions follow the naive counting rule, that is $f_L\sim 1-O(m^2_V/m^2_B), f_{\parallel}\sim f_{\perp}\sim O(m_V^2/m_B^2)$.
In the tree-dominated decay modes, such as $B^0\to \rho^+\rho^-$, where the $f_L$ is more than $90\%$. But if the contribution from the factorizable emission
amplitudes is suppressed for some decay modes, this counting
rule might be modified in some extent even dramatically
by other contributions. For example, the polarization fractions of the decay $B\to \phi K^*$ are modified by its annihilation contribution.
Whether the similar situation also occurs in the $B\to AV$ decay modes
is worth researching by theories and experiments. We know that $a_1(1260)$ has some similar behaves with the vector meson,
so one can expect that there should exist some similar characters in the branching ratios and the
polarization fractions between decays $B\to a_1(1260)V$ and $B\to \rho V$, where
$a_1(1260)$ is replaced by its scalar partner $\rho$. While it is not the case for $b_1(1235)$ because of its different characters
in decay constant and light-cone distribution amplitude (LCDA) compared with those of $a_1(1260)$. For example, the longitude decay constant is very small for the charged
$b_1(1235)$ states and vanishes under the SU(3) limit. It is zero for the neutral $b^0_1(1235)$ state. While the transverse decay constant of $a_1(1260)$ vanishes under
the SU(3) limit. In the isospin limit, the chiral-odd (-even) LCDAs of meson $b_1(1235)$ are symmetric (antisymmetric) under the exchange of quark and anti-quark momentum
fractions. It is just contrary to the symmetric behavior for $a_1(1260)$. In view of these differences, one can expect that there should exist very different
results between $B\to a_1(1260)V$ and $B\to b_1(1235)V$. On the experimental side, a few of $B\to AV$ decays are studied, such as $B\to J/\psi K_1(1270)$ \cite{kabe},
$B^0\to D^{*-}a_1^+$ \cite{babar},
$B^0\to a_1\rho$ \cite{babar2}, $B\to b_1\rho, b_1K^*$ \cite{babar3}. In most of them only the upper limits for the branching ratios can be available. On the
theoretical side, many charmless $B\to AV$ decays have been studied by Cheng and Yang in Ref. \cite{cheng} where the branching ratios are very different with
those calculated by naive factorization approach \cite{CMV}. In most cases, the former are more large than the later. To clarify such large differences is
another motivation of this work.

In the following, $a_1(1260)$ and $b_1(1235)$ are denoted as $a_1$ and $b_1$ in some places for
convenience. The layout of this paper is as follows. In Sec.\ref{proper}, decay constants and light-cone
distribution amplitudes of the relevant mesons are introduced. In Sec.\ref{results}, we then analyze these decay channels using the PQCD
approach. The numerical results and the discussions are given in
Sec. \ref{numer}. The conclusions are presented in the final part.


\section{decay constants and distribution amplitudes }\label{proper}

For the wave function of the heavy B meson,
we take
\be
\Phi_B(x,b)=
\frac{1}{\sqrt{2N_c}} (\psl_B +m_B) \gamma_5 \phi_B (x,b).
\label{bmeson}
\en
Here only the contribution of Lorentz structure $\phi_B (x,b)$ is taken into account, since the contribution
of the second Lorentz structure $\bar \phi_B$ is numerically small \cite{cdlu} and has been neglected. For the
distribution amplitude $\phi_B(x,b)$ in Eq.(\ref{bmeson}), we adopt the following model:
\be
\phi_B(x,b)=N_Bx^2(1-x)^2\exp[-\frac{M^2_Bx^2}{2\omega^2_b}-\frac{1}{2}(\omega_bb)^2],
\en
where $\omega_b$ is a free parameter, we take $\omega_b=0.4\pm0.04$ Gev in numerical calculations, and $N_B=91.745$
is the normalization factor for $\omega_b=0.4$.

The wave function for the pseudoscalar meson $P$, such as $K, \pi,
\eta^{(\prime)}$ meson is given as \be \Phi_{P}(P,x,\zeta)\equiv
\frac{1}{\sqrt{2N_C}}\gamma_5 \left [ \psl \Phi^{A}(x)+m_0
\Phi^{P}(x)+\zeta m_0 (\vsl \nsl - v\cdot n)\Phi^{T}(x)\right ].\en
where$P$ and $x$ are the momentum and the momentum fraction of the
pseudoscalar meson, respectively. The parameter $\zeta$ is either
$+1$ or $-1$ depending on the assignment of the momentum fraction
$x$.

In these decays, both the longitudinal and the transverse polarizations are involved for each final meson. For the vector mesons, their
distribution amplitudes are defined as
\be
\langle V(P, \epsilon^*_L)|\bar q_{2\beta}(z)q_{1\alpha}(0)|0\rangle&=&\frac{1}{\sqrt{2N_c}}\int^1_0dx \; e^{ixp\cdot z}[m_V\esl^*_L\phi_V(x)+\esl^*_L \psl\phi_V^{t}(x)
+m_V\phi^{s}_V(x)]_{\alpha\beta},\non
\langle V(P, \epsilon^*_T)|\bar q_{2\beta}(z)q_{1\alpha}(0)|0\rangle&=&\frac{1}{\sqrt{2N_c}}\int^1_0dx \; e^{ixp\cdot z}\left[m_V\esl^*_T\phi^v_V(x)+\esl^*_T \psl\phi_V^{T}(x)
\right.\non && \left.+m_Vi\epsilon_{\mu\nu\rho\sigma}\gamma_5\gamma^\mu\epsilon^{*v}_Tn^\rho v^\sigma\phi^{a}_V(x)\right]_{\alpha\beta},
\en
where $n (v)$ is the unit vector having the same (opposite) direction with the moving of the vector meson and  $x$ is the momentum fraction of
$q_2$ quark. The distribution amplitudes of the axial-vectors have the same format as those of the vectors except the factor $i\gamma_5$ from the left hand:
\be
\langle A(P, \epsilon^*_L)|\bar q_{2\beta}(z)q_{1\alpha}(0)|0\rangle&=&\frac{i\gamma_5}{\sqrt{2N_c}}\int^1_0dx \; e^{ixp\cdot z}[m_A\esl^*_L\phi_A(x)+\esl^*_L \psl\phi_A^{t}(x)
+m_A\phi^{s}_A(x)]_{\alpha\beta},\non
\langle A(P, \epsilon^*_T)|\bar q_{2\beta}(z)q_{1\alpha}(0)|0\rangle&=&\frac{i\gamma_5}{\sqrt{2N_c}}\int^1_0dx \; e^{ixp\cdot z}\left[m_A\esl^*_T\phi^v_A(x)+\esl^*_T \psl\phi_A^{T}(x)
\right.\non && \left.+m_Ai\epsilon_{\mu\nu\rho\sigma}\gamma_5\gamma^\mu\epsilon^{*v}_Tn^\rho v^\sigma\phi^{a}_A(x)\right]_{\alpha\beta}.
\en
\begin{table}
\caption{Decay constants and Gegenbauer moments for each meson (in MeV). The values are taken at $\mu=1$ GeV.}
\begin{center}
\begin{tabular}{|c|c|c|c|c|c|c|c|}
\hline \hline  $f_\rho$ & $f^T_\rho$ & $f_\omega$& $f^T_\omega$&$f_\phi$&$f^T_\phi$&$f_{a_1}$&$f^T_{b_1}$ \\
 $209\pm2$ &$165\pm9$&$195\pm3$&$151\pm9$&$231\pm4$&$186\pm9$&$238\pm10$ &$-180\pm8$\\
\hline
$a^\parallel_2(\rho,\omega)$&$a^\perp_2(\rho,\omega)$&$a^\parallel_2(\phi)$&$a^\perp_2(\phi)$&$a^\parallel_2(a_1(1260))$&$a^\perp_1(a_1(1260))$&$a^\parallel_1(b_1(1235))$&$a^\perp_2(b_1(1235))$\\
$0.15\pm0.07$&$0.14\pm0.06$&$0.18\pm0.08$&$0.14\pm0.07$&$-0.02\pm0.02$&$-1.04\pm0.34$&$-1.95\pm0.35$&$0.03\pm0.19$\\
\hline\hline
\end{tabular}\label{gegen}
\end{center}
\end{table}
As for the upper twist-2 and twist-3 distribution functions of the final state mesons, $\phi_{V(A)}$, $\phi_{V(A)}^t$, $\phi_{V(A)}^s$, $\phi^T_{V(A)}$,
$\phi^v_{V(A)}$ and $\phi^a_{V(A)}$ can be calculated by using light-cone QCD sum rule. We list the distribution functions of the vector (V) mesons, namely
$\rho(\omega, \phi)$, as follows
\be
\begin{cases}
\phi_V(x)=\frac{f_{V}}{2\sqrt{2N_c}}\phi_\parallel(x), \phi^T_V(x)=\frac{f^T_{V}}{2\sqrt{2N_c}}\phi_\perp(x),\\
\phi^t_V(x)=\frac{f^T_V}{2\sqrt{2N_c}}h^{(t)}_\parallel(x), \phi^s_V(x)=\frac{f^T_V}{2\sqrt{4N_c}}\frac{d}{dx}h^{(s)}_\parallel(x),\\
\phi^v_V(x)=\frac{f_V}{2\sqrt{2N_c}}g^{(v)}_\perp(x), \phi^a_V(x)=\frac{f_V}{8\sqrt{2N_c}}\frac{d}{dx}g^{(a)}_\perp(x). \label{vamp}
\end{cases}
\en
The axial-vector (A) mesons , here $a_1$ and $b_1$, can be obtain by replacing all the $\phi_V$ with $\phi_A$, by replacing
$f^T_V(f_V)$ with $f$ in Eq.(\ref{vamp}). Here we use $f$ to present both longitudinally and transversely polarized mesons $a_1(b_1)$ by assuming $f^T_{a_1}=f_{a_1}=f$ for $a_1$
and $f_{b_1}=f^T_{b_1}=f$ for $b_1$. In Eq.(\ref{vamp}), the twist-2 distribution functions are in the first line and can be expanded as
\be
\phi_{\parallel,\perp}&=&6x(1-x)\left[1+a^{\parallel,\perp}_2\frac{3}{2}(5t^2-1)\right], \quad\quad\quad\quad\quad\quad\mbox{ for $V$ mesons};\label{t1} \\
\phi_{\parallel,\perp}&=&6x(1-x)\left[a^{\parallel,\perp}_0+3a^{\parallel,\perp}_1t+a^{\parallel,\perp}_2\frac{3}{2}(5t^2-1)\right], \quad\mbox{ for $A$ mesons},
\en
where the zeroth Gegenbauer moments $a^{\perp}_0(a_1)=a^{\parallel}_0(b_1)=0$ and $a^{\parallel}_0(a_1)=a^{\perp}_0(b_1)=1$.

As for twist-3 LCDAs, we use the asymptotic forms for $V$ mesons:
\be
h^{(t)}_\parallel(x)&=&3t^2, h^{(s)}_\parallel(x)=6x(1-x),\non
g^{(a)}_\perp(x)&=&6x(1-x), g^{(v)}_\perp(x)=\frac{3}{4}(1+t^2).
\en
And we use the following forms for $A$ mesons:
\be
h^{(t)}_\parallel(x)&=&3a^\perp_0t^2+\frac{3}{2}a^\perp_1t(3t^2-1), h^{(s)}_\parallel(x)=6x(1-x)(a^\perp_0+a^\perp_1t),\non
g^{(a)}_\perp(x)&=&6x(1-x)(a^\parallel_0+a^\parallel_1t), g^{(v)}_\perp(x)=\frac{3}{4}a^\parallel_0(1+t^2)+\frac{3}{2}a^\parallel_1t^3. \label{t4}
\en
In Eqs.(\ref{t1})-(\ref{t4}), the function $t=2x-1$. As in Ref.\cite{rhli}, the decays constants and the Gegenbauer moments $a^{\parallel,\perp}_n$ for each meson are quoted the numerical
results \cite{camsler, pball1, pball2, pball3, yang1, yang2} and listed in Table \ref{gegen}.


\section{ the perturbative QCD  calculation} \label{results}
The PQCD approach is an effective theory to handle hadronic $B$ decays \cite{cdlu2,keum,mishima}. Because it takes into account the transverse momentum of the valence
quarks in the hadrons, one will encounter double logarithm divergences when the soft and the collinear momenta overlap. Fortunately, these large
double logarithm can be re-summed into the Sudakov factor \cite{hnli0}. There are also another type of double logarithms which arise from the loop corrections
to the weak decay vertex. These double logarithms can also be re-summed and resulted in the threshold factor \cite{hnli00}. This factor decreases faster than any other
power of the momentum fraction in the threshold region, which removes the endpoint singularity. This factor is often parameterized
into a simple form which is independent on channels, twists and flavors \cite{hnli}. Certainly, when the higher order diagrams only suffer from soft or
collinear infrared divergence, it is ease to cure by using the eikonal approximation \cite{hnli2}. Controlling these
kinds of divergences reasonably makes the PQCD approach more self-consistent.

\begin{figure}[t,b]
\vspace{-3cm} \centerline{\epsfxsize=16 cm \epsffile{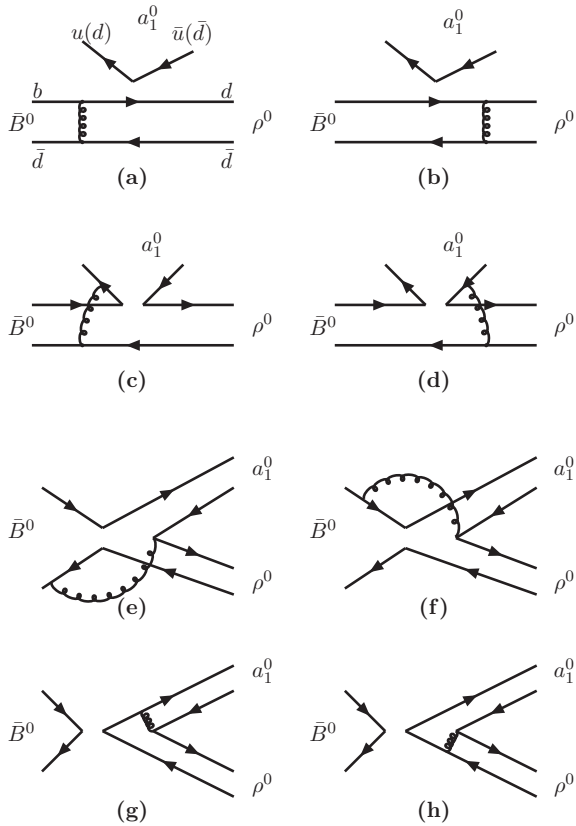}}
\vspace{-9cm} \caption{ Diagrams contributing to the decay $\bar{B}^0\to a^0_1\rho^0$.}
 \label{fig1}
\end{figure}

Here we take the decay $\bar{B}^0\to a^0_1\rho^0$ as an example, whose part of diagrams are shown in Figure 1. These eight Feynman diagrams belong to the condition
of $a_1^0$ meson being at the emission position. Another eight Feynman diagrams obtained by exchanging the positions of $a_1^0$ and $\rho^0$ in Fig.1 also
contribute to the decay. All of these single hard gluon exchange diagrams contain all of the leading order contributions to $\bar{B}^0\to a^0_1\rho^0$
in the PQCD approach. Similar to the $B\to VV$
decay modes, such as $B\to \rho\rho$ \cite{yli}
and $B\to K^*\rho(\omega)$ \cite{hwhuang}, both longitudinal and transverse polarizations can contribute to the decay width. So we can get three kinds of polarization
amplitudes $M_L$ (longitudinal) and $M_{N,T}$ (transverse) by calculating these diagrams. Because of the aforementioned distribution amplitudes of the axial-vectors
having the same format as those of the vectors except a factor, so the formulas of here considered decays can be obtained from the ones
of $B\to VV$ decays by some replacements. Certainly, there also
exists a difference: if the emitted meson is $b_1$ for the factorizable emission diagrams, the amplitudes contributed by the $(V-A)(V\pm A)$ operators would be zero due to the
vanishing decay constant $f_{b_1}$. For the tree-dominated decays, the contributions from the factorizable emission diagrams, namely Fig.(a),(b), are very
important. In the PQCD approach, the form factor can be extracted from the amplitudes obtained by calculating such diagrams, where
the two transverse amplitudes are highly suppressed by the factor $r_{a_1(b_1)}\cdot r_{\rho(\omega)}$ compared with the longitudinal amplitudes. Here $r_{a_1(b_1)}=\frac{m_{a_1(b_1)}}{m_B}$ and
$r_{\rho(\omega)}=\frac{m_{\rho(\omega)}}{m_B}$. To some decays, the non-factorizable emission diagrams, namely Fig.(c),(d),
play an more important role, where the contributions from the transverse polarizations are not suppressed. Certainly, the
contributions from the non-factorizable and the factorizable annihilation diagrams, that are Fig.(g),(h) and Fig.(e),(f), can also not be neglected.
\section{Numerical results and discussions} \label{numer}

We use the following input parameters in the numerical calculations \cite{pdg10,ckmfit}:
\be
f_B&=&190 MeV, M_B=5.28 GeV, M_W=80.41 GeV,\\
\tau_{B^\pm}&=&1.638\times 10^{-12} s,\tau_{B^0}=1.525\times 10^{-12} s,\\
|V_{ud}|&=&0.974,  |V_{td}|=8.58\times10^{-3}, \alpha=(91.0\pm3.9)^\circ,\\
|V_{ub}|&=&3.54\times10^{-3}, |V_{tb}|=0.999.
\en

In the B-rest frame, the decay rates of $B\to a_1(b_1)V$, where $V$ represents $\rho, \omega, \phi$,
can be written as
\be \Gamma=\frac{G_F^2(1-r^2_{a_1(b_1)})}{32\pi M_B}\sum_{\sigma=L,N,T}{\cal M}^{\sigma\dagger}{\cal M}^{\sigma}, \en
where ${\cal M}^{\sigma}$ is the total decay amplitude
of each considered decay. The subscript $\sigma$ is the helicity states of the two final
mesons with one longitudinal component and two transverse ones. The decay amplitude can be decomposed into three scalar amplitudes $a, b, c$
according to
\be
{\cal M}^{\sigma}&=&\epsilon^*_{2\mu}(\sigma)\epsilon^*_{3\nu}(\sigma)\left[a g^{\mu\nu}+\frac{b}{M_2M_3}P^\mu_BP^\nu_B+
i\frac{c}{M_2M_3}\epsilon^{\mu\nu\alpha\beta} P_{2\alpha}P_{3\beta}\right]\non
&=&{\cal M}_L+{\cal M}_N\epsilon^*_2(\sigma=T)\cdot\epsilon^*_3(\sigma=T)+i\frac{{\cal M}_T}{M^2_B}\epsilon^{\alpha\beta\gamma\rho}
\epsilon^*_{2\alpha}(\sigma)\epsilon^*_{3\beta}(\sigma)P_{2\gamma}P_{3\rho},
\en
where $M_2$ and $M_3$ are the masses of the two final mesons $a_1(b_1)$ and $\rho(\omega,\phi)$, respectively. The amplitudes
${\cal M}_L, {\cal M}_N, {\cal M}_T$ can be
expressed as
\be
{\cal M}_L&=&a \;\epsilon^*_2(L)\cdot\epsilon^*_3(L)+\frac{b}{M_2M_3}\epsilon^*_{2}(L)\cdot P_3\epsilon^*_{3}(L)\cdot P_2,\non
{\cal M}_N&=&a, \;\;\;\;{\cal M}_T=\frac{M^2_B}{M_2M_3}c.
\en
We can use the amplitudes with different Lorentz structures to define the helicity amplitudes, one longitudinal amplitudes $H_0$
and two transverse amplitudes $H_{\pm}$:
\be
H_0=M^2_B{\cal M}_L,\;\;\; H_{\pm}=M^2_B{\cal M}_N\mp M_2M_3\sqrt{r^2-1}{\cal M}_T,
\label{ampde} \en
where the ratio $r=P_2\cdot P_3/(M_2M_3)$. After the helicity summation, we can get the relation
\be
\sum_{\sigma=L,N,T}{\cal M}^{\sigma\dag}{\cal M}^\sigma=|{\cal M}_L|^2+2\left(|{\cal M}_N|^2+|{\cal M}_T|^2\right)=|H_0|^2+|H_+|^2+|H_-|^2.
\en

Certainly another equivalent set of helicity amplitudes are often used, that is
\be
A_0&=&-M^2_B{\cal M}_L,\non
A_{\parallel}&=&\sqrt{2}M^2_B{\cal M}_N,\non
A_{\perp}&=&M_2M_3\sqrt{2(r^2-1)}{\cal M}_T.
\en
Using this set of helicity amplitudes, we can define three polarization fractions $f_{0, \parallel, \perp}$:
\be
f_{0,\parallel,\perp}=\frac{|A_{0,\parallel,\perp}|^2}{|A_0|^2+|A_\parallel|^2+|A_\perp|^2}.
\en

The matrix elements ${\cal M}_{j}$ of the operators in the weak Hamilitonian can be calculated by using PQCD approach,
which are written as
as
\be
M_j&=&V_{ub}V^*_{ud}T_{j}-V_{tb}V^*_{td}P_{j}\non
&=&V_{ub}V^*_{ud}T_{j}(1+z_je^{i(\alpha+\delta_j)}), \label{am}
\en
where $j=L, N, T$ and $\alpha$ is the Cabibbo-Kobayashi-Maskawa
weak phase angle, defined via $\alpha=arg[-\frac{V_{td}V^*_{tb}}{V_{ud}V^*_{ub}}]$. Here we leave this angle as a free parameter.
$\delta_{j}$ is the relative strong phase between
the tree and the penguin amplitudes, which are denoted as "$T_{j}$" and
"$P_{j}$", respectively. The term $z_{j}$ describes the ratio of penguin to
tree contributions and is defined as \be
z_j=\left|\frac{V_{tb}V^*_{td}}{V_{ub}V^*_{ud}}\right|\left|\frac{P_j}{T_j}\right|.
\en
In the same way, it is easy to write decay amplitude
$\overline {\cal M}_j$ for the corresponding conjugated decay mode:
\be
\overline {\cal M}_j&=&V^*_{ub}V_{ud}T_{j}-V^*_{tb}V_{td}P_{j}\non
&=&V^*_{ub}V_{ud}T_{j}(1+z_je^{i(-\alpha+\delta_j)}).\label{amcon}
\en
So the CP-averaged branching ratio for each considered decay is defined
as \be {\cal B}=(|{\cal M}_j|^2+|\overline{\cal
M}_j|^2)/2&=&|V_{ub}V^*_{ud}|^2\left[T^2_L(1+2z_L\cos\alpha\cos\delta_L+z_L^2)\right.\non &&\left.
+2\sum_{j=N,T}T^2_j(1+2z_j\cos\alpha\cos\delta_j+z_j^2)\right].\label{brann}
\en
Like the decays of $B$ to two vector mesons, there are also $3$ types of helicity amplitudes, so corresponding to $3$ types of $z_j$ and $\delta_j$,
respectively. It is easy to see that the dependence of decay width on $\delta$ and $\alpha$ is more complicated compared with that for the
decays of $B$ to pseudoscalar mesons.

\begin{figure}[t,b]
\begin{center}
\includegraphics[scale=0.7]{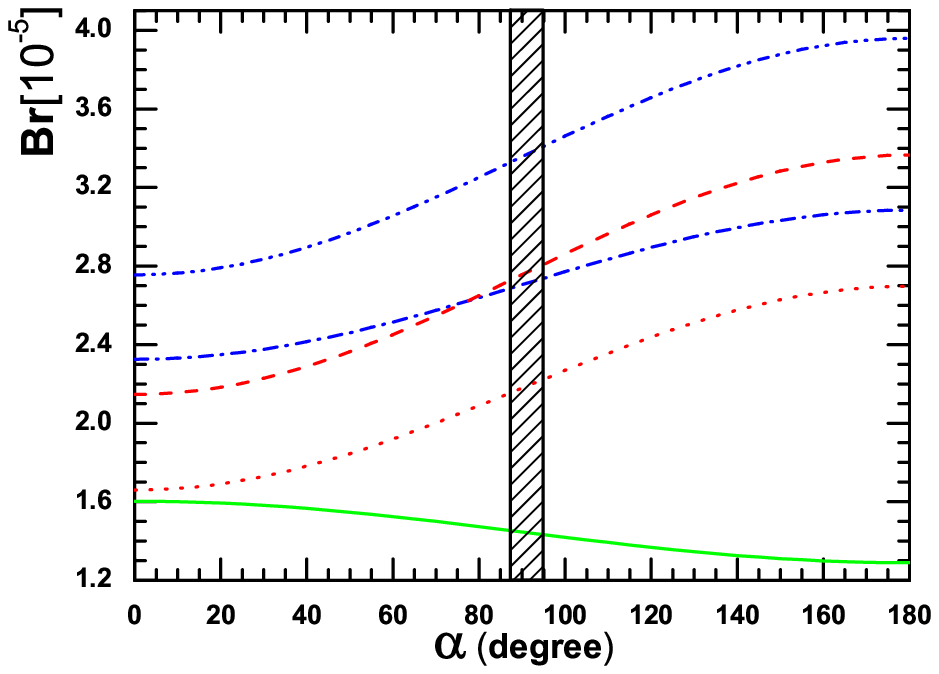}
\includegraphics[scale=0.7]{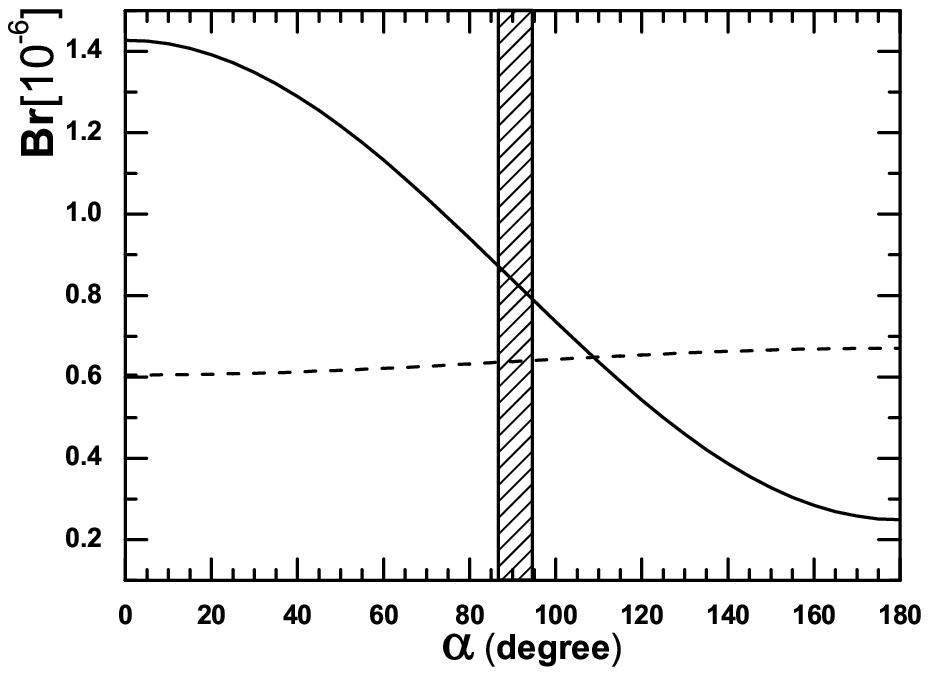}
\vspace{0.3cm} \caption{The dependence of the branching ratios on the
Cabibbo-Kobayashi-Maskawa angle $\alpha$. In the left panel, the solid line is for $B^-\to  a^{-}_1\omega$, dotted line for $B^-\to  a^{-}_1\rho^0$,
dot-dashed line for $\bar B^0\to  a^{-}_1\rho^+$, dashed line for $B^-\to  b^{0}_1\rho^-$, dot-dot-dashed line for $\bar B^0\to  a^{+}_1\rho^-$.
In the right panel, the solid line is for $\bar B^0\to  a^{0}_1\omega$ and the dashed line is for $\bar B^0\to  a^{0}_1\rho^0$.
The vertical band shows the range of $\alpha$: $91.0\pm3.9$.}\label{fig2}
\end{center}
\end{figure}

\begin{figure}[t,b]
\begin{center}
\includegraphics[scale=0.7]{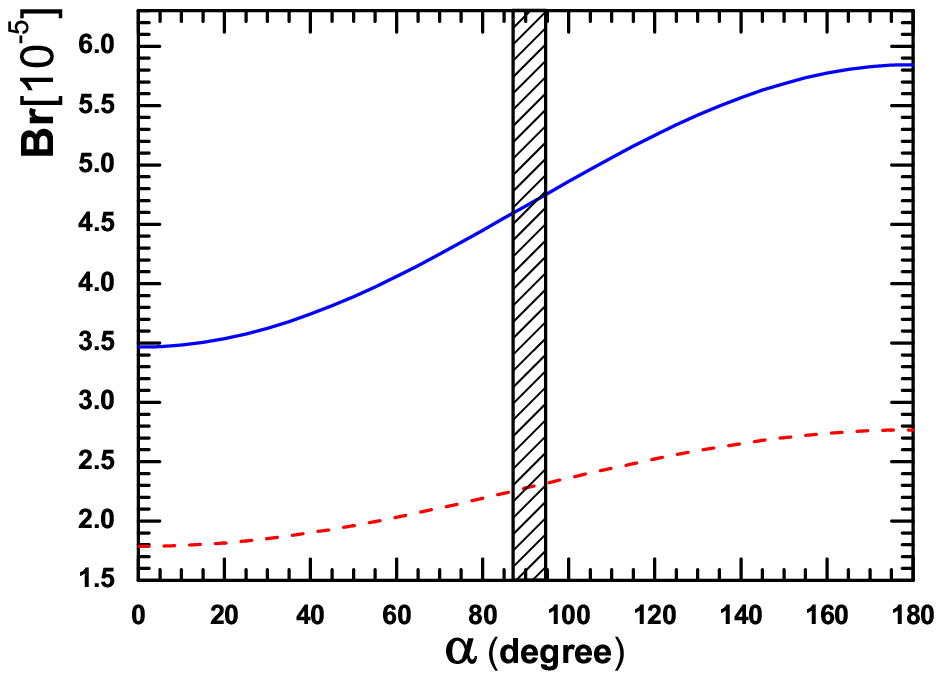}
\includegraphics[scale=0.7]{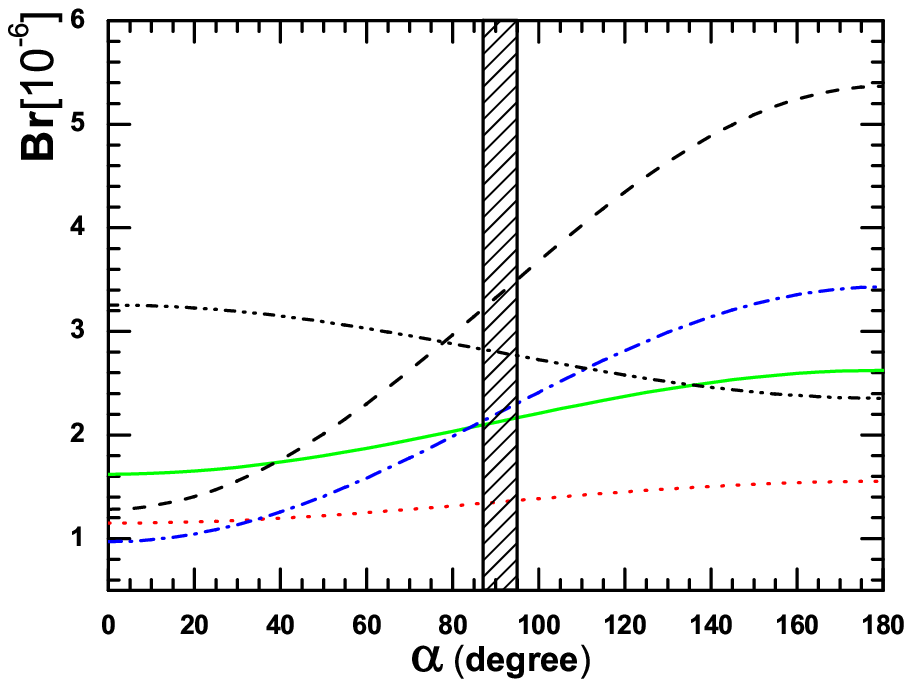}
\vspace{0.3cm} \caption{The dependence of the branching ratios on the
Cabibbo-Kobayashi-Maskawa angle $\alpha$. In the left panel, the solid line is for $\bar B^0\to  b^{+}_1\rho^-$
and the dashed line is for $B^-\to  b^{0}_1\rho^-$. In the right panel, the dotted line is for $B^-\to  b^{-}_1\rho^0$, solid line for $B^-\to  b^{-}_1\omega$,
dot-dashed line for $\bar B^0\to  b^{-}_1\rho^+$, dot-dot-dashed line for $\bar B^0\to  b^{0}_1\omega$ and dashed line for $\bar B^0\to  b^{0}_1\rho^0$.
The vertical band shows the range of $\alpha$: $91.0\pm3.9$.}\label{fig3}
\end{center}
\end{figure}

Using the input parameters and the wave functions as specified in this section and Sec.\ref{proper}, it is easy to get the branching ratios for
the considered decays which are listed in Table \ref{bran},
\begin{table}
\caption{ Branching ratios (in units of $10^{-6}$) for the decays $B\to a_1(1260)\rho(\omega, \phi)$ and
$B\to b_1(1235)\rho(\omega, \phi)$. In our results, the errors for these entries correspond to the uncertainties from $\omega_B$ and threshold resummation
parameter $c$, respectively. For comparison, we also listed the results predicted by QCDF approach \cite{cheng} and naive factorization approach
\cite{CMV}.}
\begin{center}
\begin{tabular}{c|c|c|c|c|c}
\hline\hline   & This work  & \cite{cheng} & \cite{CMV}\\
\hline
$\bar B^0\to  a^{+}_1\rho^-$ &$33.6^{+9.9+15.8}_{-7.4-15.8}$&$23.9^{+10.5+3.2}_{-9.2-0.4}$&$4.3$\\
$\bar B^0\to  a^{-}_1\rho^+$ &$27.1^{+8.0+9.2}_{-6.0-9.2}$&$36.0^{+3.5+3.5}_{-4.0-0.7} $&$4.7$\\
$\bar B^0\to  a^{0}_1\rho^0$ &$0.64^{+0.12+0.04}_{-0.10-0.04}$&$1.2^{+2.0+5.1}_{-0.7-0.3} $&$0.01$\\
$B^-\to  a^{0}_1\rho^-$&$27.7^{+7.8+7.9}_{-5.9-7.9}$&$17.8^{+10.1+3.1}_{-6.4-0.2} $&$2.4$\\
$B^-\to  a^{-}_1\rho^0$ &$21.9^{+5.9+9.3}_{-4.6-9.3}$&$23.2^{+3.6+4.8}_{-2.9-0.1} $&$3.0$\\
$\bar B^0\to  a^{0}_1\omega$ &$0.83^{+0.27+0.40}_{-0.20-0.40}$&$0.2^{+0.1+0.4}_{-0.1-0.0} $&$0.003$\\
$B^-\to  a^{-}_1\omega$ &$14.4^{+4.8+6.0}_{-3.5-6.0}$&$22.5^{+3.4+3.0}_{-2.7-0.7}$&$2.2$\\
$\bar B^0\to  a^{0}_1\phi$ &$0.0029^{+0.0007+0.0006}_{-0.0006-0.0006}$&$0.002^{+0.002+0.009}_{-0.001-0.000} $&$0.0005$\\
$B^-\to  a^{-}_1\phi$ &$0.0058^{+0.0015+0.0013}_{-0.0013-0.0013}$&$0.01^{+0.01+0.04}_{-0.00-0.00} $&$0.001$\\
\hline
$\bar B^0\to  b^{+}_1\rho^-$ &$46.8^{+15.6+19.1}_{-11.3-19.1}$&$32.1^{+16.5+12.0}_{-14.7-4.6} $&$1.6$\\
$\bar B^0\to  b^{-}_1\rho^+$ &$2.2^{+0.3+0.1}_{-0.3-0.1}$&$0.6^{+0.6+1.9}_{-0.3-0.2} $&$0.55$\\
$\bar B^0\to  b^{0}_1\rho^0$ &$3.4^{+0.4+0.4}_{-0.5-0.4}$&$3.2^{+5.2+1.7}_{-2.0-0.4} $&$0.002$\\
$B^-\to  b^{0}_1\rho^-$&$22.9^{+8.7+24.3}_{-6.3-24.3}$&$29.1^{+16.2+5.4}_{-10.6-5.9} $&$0.86$\\
$B^-\to  b^{-}_1\rho^0$ &$1.4^{+0.2+0.3}_{-0.2-0.3}$&$0.9^{+1.7+2.6}_{-0.6-0.5} $&$0.36$\\
$\bar B^0\to  b^{0}_1\omega$ &$2.8^{+0.7+0.2}_{-0.6-0.2}$&$0.1^{+0.2+1.6}_{-0.0-0.0} $&$0.004$\\
$B^-\to  b^{-}_1\omega$ &$2.1^{+0.4+0.7}_{-0.2-0.7}$&$0.8^{+1.4+3.1}_{-0.5-0.3} $&$0.38$\\
$\bar B^0\to  b^{0}_1\phi$ &$0.003^{+0.001+0.000}_{-0.001-0.000}$&$0.01^{+0.01+0.01}_{-0.00-0.00} $&$0.0002$\\
$B^-\to  b^{-}_1\phi$ &$0.006^{+0.003+0.001}_{-0.002-0.001}$&$0.02^{+0.02+0.03}_{-0.01-0.00}$&$0.0004$\\
\hline\hline
\end{tabular}\label{bran}
\end{center}
\end{table}
where the first error comes from the uncertainty in the $B$ meson shape parameter $\omega_b=0.40\pm0.04$ GeV, the second one is from the
threshold resummation parameter $c$, and it varies from $0.3$ to $0.4$. In Fig.\ref{fig2} and Fig.\ref{fig3}, we also show the Cabibbo-Kobayashi-Maskawa angle $\alpha$
dependence of the branching ratios of decays $B\to a_1\rho(\omega)$ and $B\to b_1\rho(\omega)$.


From Table \ref{bran}, one can find that except decays $\bar B^0\to  a^{0}_1\rho^0, a^0_1\omega$, the branching ratios of other tree-dominated
decays $B\to a_1 \rho(\omega)$ are all at the order of $10^{-5}$. Most of the contributions to such larger branching ratios are from the
factorizable
emission diagrams $(a)$ and $(b)$, which contribute to the $B\to \rho(\omega)$ ($B\to a_1$) form factors. Because of the large Wilson coefficients
$C_2+C_1/3$
in the amplitudes contributed by the tree operators $O_1$ and $O_2$, the branch ratios are almost proportionate to the corresponding form
factors. Certainly,
they are also related to the decay constants $f_{a_1}$ ($f_{\rho,\omega}$). As the basic input values, they are the same in many factorization approaches, for example,
PQCD and QCDF approaches. While for the form factors, there
exist some differences between these two approaches. For QCDF approach, the form factors are used as the input values, which are obtained
from light-cone sum rules. In Ref.
\cite{cheng}, the form factors $A^{B\rho}_0$ and $V^{Ba_1}_0$ are both about $0.30$, and $V^{Bb_1}_0$ is about $-0.39$, where the authors put an additional minus
sign by taking the convention of the decay constants of $a_1$ and $b_1$ being of the same sign. In this convention, the corresponding form factors have opposite signs.
For the PQCD approach, the form factors can be calculated perturbatively. From our calculations, we find that the values of $A^{B\rho}_0$, $V^{Ba_1}_0$
and $V^{Bb_1}_0$ are about $0.25$, $0.33$ and $0.44$, respectively. If the decay is governed by the form factor $A^{B\rho}_0$, its branching ratio predicted by PQCD approach would be
smaller than that obtained by QCDF approach, for example, $\bar B^0\to  a^{-}_1\rho^+$. On the contrary, if the decay is governed by the form factor $V^{Ba_1}_0$, the
result for the PQCD approach would have a larger value, the decay $\bar B^0\to  a^{+}_1\rho^-$ is in this case. So to accurately determine these form factors is
very important. The branching ratio of $B^-\to  a^{-}_1\rho^0$ is larger than that of $B^-\to  a^{-}_1\omega$, one
reason is that the form factor $A^{B\rho}_0$ is a litter larger than $A^{B\omega}_0$, which is about 0.23. The other reason is the different interferences from
$d\bar d$ and $u\bar u$: constructive interference between $-d\bar d$ and $u\bar u$ which compose $\rho$, destructive interference between $d\bar d$ and $u\bar u$ which compose $\omega$.
But there is a contrary situation for the QCDF approach between these two decays. Although the neutral decays
$\bar B^0\to  a^{0}_1\rho^0, a^0_1\omega$ are also tree
dominant, their tree operator contributions are highly suppressed compared with the two charged decays $B^-\to  a^{-}_1\rho^0, a^{-}_1\omega$ (shown in
Table \ref{amppp}). So their branching ratios are small
and at the order of $10^{-7}$.
Certainly, we only give the leading order results and they might like decays $B\to \rho^0\rho^0, \rho^0\omega$, which are sensitive to the next leading order
contributions.
\begin{table}
\caption{ Polarization amplitudes of different diagrams for the decays $\bar B^0\to a^{+}_1\rho^{-},
a^{0}_1\rho^{0}$ ($\times 10^{-2} \mbox {GeV}^3$).}
\begin{center}
\begin{tabular}{c|c|c|c|c|c|c|c|c|c|c}
\hline \hline  Decay mode &Pol. amp. & (a) and (b) & (c) and (d) &(e) and (f)& (g) and (h) \\
\hline
  &$A(T_L)$&-219.2&$8.1-3.8i$&$-1.2+9.0i$&$-0.5-0.1i$ \\
  &$A(T_N)$&22.8&$-7.1+5.2i$&$-0.2-0.2ii$&$0.6+0.03i$ \\
$ a^{+}_1\rho^{-}$  &$A(T_T)$&-57.3&$-12.9+3.2i$&$0.1-0.4i$&$-1.0-0.2i$ \\
   &$A(P_L)$&8.8&$-0.09+0.13i$&$0.59+1.7i$&$-1.7-3.4i$\\
   &$A(P_N)$&0.9&$0.26-0.17i$&$-0.03-0.01i$&$0.7+3.4i$\\
   &$A(P_T)$&2.2&$0.49-0.05i$&$-0.03+0.01i$&$1.1+6.6i$\\
\hline
  &$A(T_L)$&-5.7&$18.4-7.3i$&$1.1-4.7i$&$1.4-1.0i$ \\
  &$A(T_N)$&0.5&$-11.0+6.5i$&$0.06+0.06i$&$0.54+0.05i$ \\
$ a^{0}_1\rho^{0}$  &$A(T_T)$&-0.1&$-20.8+5.2i$&$0.00+0.17i$&$-1.1-0.15i$ \\
  &$A(P_L)$&0.8&$0.36+0.12i$&$0.33+1.24i$&$-0.12-0.06i$\\
   &$A(P_N)$&-0.15&$0.26-0.15i$&$-0.05-0.02i$&$-0.08+0.04i$\\
   &$A(P_T)$&-0.06&$0.50-0.1i$&$-0.08+0.00i$&$-0.34-0.14i$\\
\hline\hline
\end{tabular}\label{amppp}
\end{center}
\end{table}

As to the tree-dominated decays $B\to b^+_1\rho^-, b^0_1\rho^-$, which are governed by the decay constant $f_{\rho}$ and the form factor of
$B\to b_1$, they also have large branching ratios. Although $B\to b^-_1\rho^+$ is color allowed decay, its branching ratio is highly suppressed due to the
decay constant $f_{b_1}$ being
very small and vanishing under the isospin limit.
One should admit that each amplitude for the decays $B^-\to b_1^-(b_1^0)\rho^0$ has near value in magnitude with the corresponding one for
the decays $B^-\to b_1^-(b_1^0)\omega$, but the sign differences before $d\bar d$ in the mesons $\rho$ and $\omega$ will induce some discrepancies in the
branching ratios.
Like the decays $B\to \pi\phi, a_0(1450)\phi$ \cite{yli2,zqzhang}, whose branching ratios are at the order of $10^{-8}\sim 10^{-9}$, the decays $B\to a_1(b_1)\phi$ are
induced by the flavor-changing neutral current (FCNC) interactions and highly suppressed by the small Wilson coefficients for penguin operators. Moreover, there
is no the contribution from the annihilation diagram. So one expects that their branching ratios are also very small.

From Table \ref{bran}, One can find that our predictions are well consistent with the results calculated by QCDF approach for most decays. Certainly, there also exist large differences
for some decays, which are needed to clarify by the present LHCb experiments. At the present, BaBar has given the upper limits of the branching ratios for the decays
$B\to b_1\rho$, ranging from $1.4\sim 5.2\times 10^{-6}$ at the $90\%$ confidence level \cite{babar3}, which are not far away from our predictions for
the decays $\bar B^0\to  b^{0}_1\rho^0$ and $B^-\to  b^{-}_1\rho^0$, but much smaller than those of $\bar B^0\to  b^{+}_1\rho^-$ and $B^-\to  b^{0}_1\rho^-$.
In Ref.\cite{babar2}, the BarBar collaboration searched the decay
$\bar B^0\to a_1^\pm\rho^\mp$ and obtained an upper limit of $61\times10^{-6}$ by assuming that $a^{\pm}_1$ decays exclusively to $\rho^0\pi^{\pm}$. Our prediction
for the branching ratio of $\bar B^0\to a_1^\pm\rho^\mp$ is about $60\times 10^{-6}$, which agrees with the experiment.

\begin{table}
\caption{Longitudinal polarization fraction ($f_L$) and two transverse polarization fractions ($f_\parallel$, $f_\perp$) for the decays
$B\to a_1(1260)\rho(\omega)$ and $B\to b_1(1235)\rho(\omega)$. In our results, the uncertainties of $f_L$ come from $\omega_{B}$ and threshold resummation
parameter $c$. The results of $f_L$ predicted by the QCDF approach are also displayed in parentheses for comparison.}
\begin{center}
\begin{tabular}{ccccccc|c}
\hline\hline   & $f_L(\%)$  &$f_\parallel(\%)$&$f_\perp(\%)$\\
\hline
$\bar B^0\to  a^{+}_1\rho^-$&$90.7^{+0.2+1.3}_{-0.2-1.3}(82^{+5}_{-13})$&$3.9$&$5.4$\\
$\bar B^0\to  a^{-}_1\rho^+$ &$90.4^{+0.0+0.1}_{-0.1-0.1} (84^{+2}_{-6})$&$5.2$&$4.4$\\
$\bar B^0\to  a^{0}_1\rho^0$ &$43.3^{+1.2+2.9}_{-1.3-2.9} (82^{+6}_{-68})$&$29.7$&$27.0$\\
$B^-\to  a^{0}_1\rho^-$&$93.6^{+0.2+0.1}_{-0.2-0.1} (91^{+3}_{-10})$&$2.8$&$3.6$\\
$B^-\to  a^{-}_1\rho^0$ &$82.3^{+0.1+2.0}_{-0.3-2.0} (89^{+11}_{-18})$&$9.3$&$8.4$\\
$\bar B^0\to  a^{0}_1\omega$ &$80.7^{+0.3+3.4}_{-0.1-3.4} (75^{+11}_{-65})$&$9.9$&$9.4$\\
$B^-\to  a^{-}_1\omega$ &$79.5^{+0.6+2.2}_{-0.6-2.2} (88^{+10}_{-14})$&$8.9$&$11.6$\\
\hline
$\bar B^0\to  b^{+}_1\rho^-$ &$95.4^{+0.2+0.1}_{-0.1-0.1} (96^{+1}_{-2})$&$2.2$&$2.4$\\
$\bar B^0\to  b^{-}_1\rho^+$ &$95.8^{+0.5+1.1}_{-0.5-1.1} (98^{+0}_{-33})$&$1.7$&$2.5$\\
$\bar B^0\to  b^{0}_1\rho^0$ &$95.3^{+0.2+0.4}_{-0.4-0.4} (99^{+0}_{-18})$&$2.8$&$1.9$\\
$B^-\to  b^{0}_1\rho^-$      &$92.5^{+0.9+0.6}_{-1.1-0.6} (96^{+1}_{-6})$&$0.8$&$6.7$\\
$B^-\to  b^{-}_1\rho^0$ &$44.9^{+1.8+5.6}_{-2.0-5.6} (90^{+5}_{-38})$&$1.1$&$54.0$\\
$\bar B^0\to  b^{0}_1\omega$ &$93.5^{+0.2+0.3}_{-0.1-0.3} (4^{+96}_{-0})$&$4.3$&$2.2$\\
$B^-\to  b^{-}_1\omega$ &$73.1^{+0.5+1.0}_{-0.6-1.0} (91^{+7}_{-33})$&$25.5$&$1.4$\\
\hline\hline
\end{tabular}\label{polar}
\end{center}
\end{table}

In Table \ref{polar}, we list the polarization fractions of $B\to a_1\rho(\omega), b_1\rho(\omega)$ decays and find that the longitudinal polarizations are dominant
in most of these decays, which occupy more than $80\%$. For the tree-dominated decays, the main contributions come from the factorizable emission diagrams, where the
two kinds of transverse polarization amplitudes are highly suppressed by the aforementioned factor $r_{a_1(b_1)}\cdot r_{\rho(\omega)}$. From Table \ref{polar}, One can find
that $f_\parallel$ and $f_\perp$ have near values and both about a few percent in general. Certainly, for the decays $\bar B^0\to  a^{0}_1\rho^0$ and
$B^-\to b^{-}_1\rho^0(\omega)$, their polarization fractions are very different with those of other decays. In the decay $\bar B^0\to  a^{0}_1\rho^0$, the contributions
from the two transverse polarization components become prominent and are larger than that from the longitudinal component. It is because that the decay is
suppressed by the cancelation of Wilson coefficients $C_1+C_2/3$ for the color-suppressed amplitude. So the contribution from
the factorizable emission diagrams become very small. The left dominant contributions are the non-factorizable amplitudes from tree operators,
where either of the transverse polarizations is not suppressed compared with the longitudinal polarization. Therefore numerically we get a small
longitudinal polarization fraction of about
$43\%$. In Table \ref{pola1}, if we ignore the contribution from the non-factorizable amplitudes of $\bar B^0\to  a^{0}_1\rho^0$ and find that the longitudinal
polarization becomes dominant, but the
branching ratio becomes very small. If we ignore the contributions from its penguin operators or annihilation diagrams, the results have small changes.
As to the other charged decays $B^-\to b^{-}_1\rho^0(\omega)$, either of their transverse polarizations is very sensitive to the contributions listed in lines (2)-(4) in Table \ref{pola1}.

\begin{table}
\caption{ Contributions from different parts in the decays $\bar B^0\to  a^{0}_1\rho^0$ and $B^-\to  b^{-}_1\omega$: line (1) is for full contribution, line (2), (3) and (4) are the contributons
after ignoring annihilation diagrams, penguin operators and non-factorization diagrams, respectively.}
\begin{center}
\begin{tabular}{|c|c|c|c|c|}
\hline\hline   $\bar B^0\to  a^{0}_1\rho^0$ &Br$(10^{-7})$& $f_L(\%)$ &$f_\parallel(\%)$&$f_\perp(\%)$\\
\hline
(1)&$6.4$&$43.3$&$29.7$&$27.0$\\
(2)&$5.1$&$28.4$&$40.4$&$31.2$\\
(3) &$6.3$&$42.5$&$30.1$&$27.4$\\
(4)&$0.2$&$86.1$&$9.4$&$4.5$\\
\hline\hline   $B^-\to  b^{-}_1\omega$ &Br$(10^{-6})$& $f_L(\%)$ &$f_\parallel(\%)$&$f_\perp(\%)$\\
\hline
(1)&$2.1$&$73.1$&$25.5$&$1.4$\\
(2)&$0.9$&$63.5$&$18.5$&$18.0$\\
(3) &$0.7$&$67.9$&$0.1$&$32.0$\\
(4)&$1.8$&$83.2$&$11.1$&$5.7$\\
\hline\hline
\end{tabular}\label{pola1}
\end{center}
\end{table}
\begin{figure}[t,b]
\begin{center}
\includegraphics[scale=0.7]{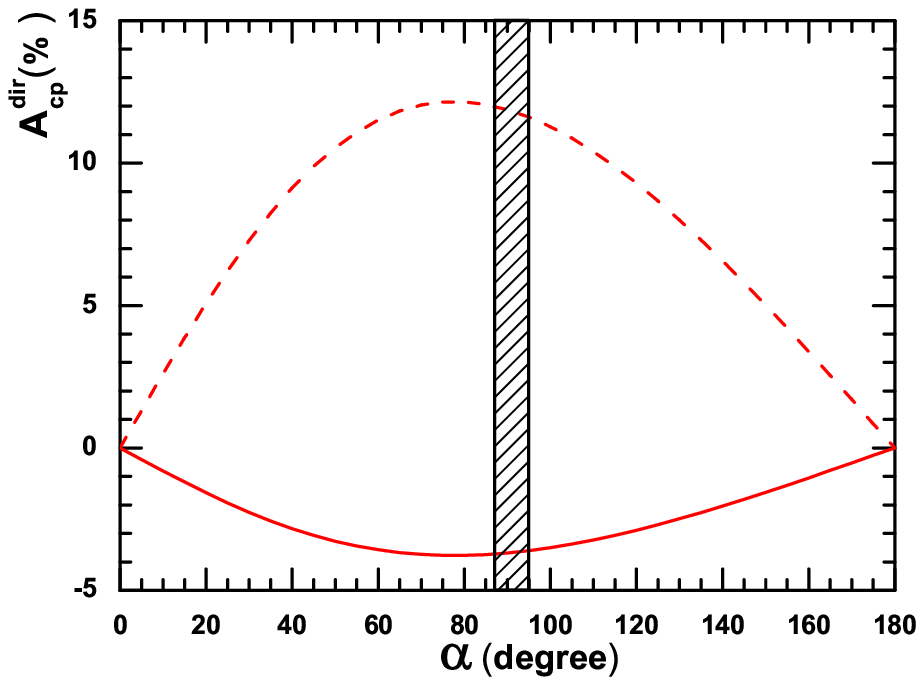}
\includegraphics[scale=0.7]{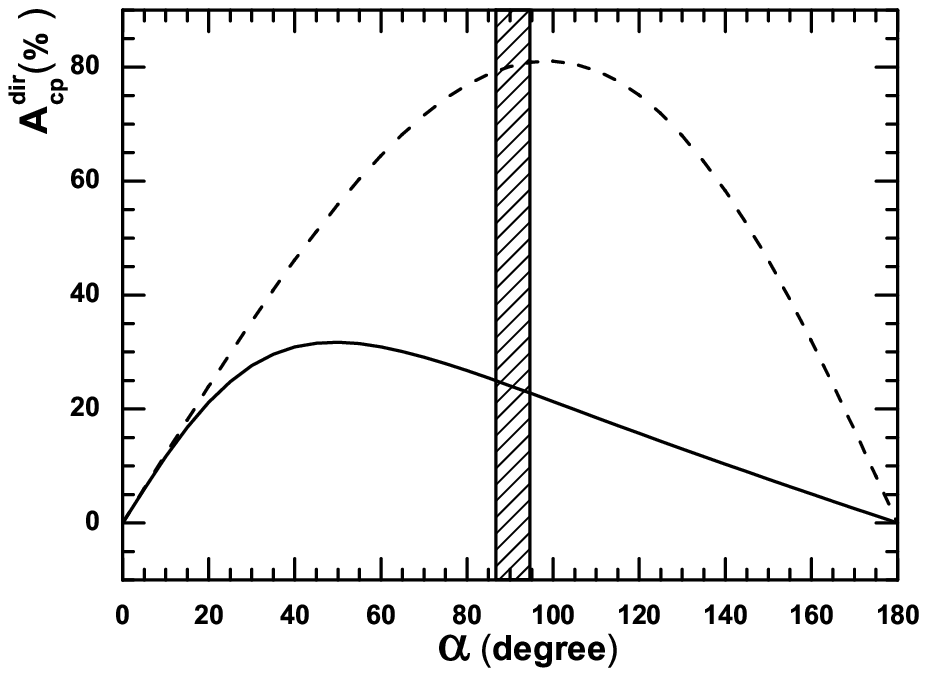}
\vspace{0.3cm} \caption{Direct CP-violating asymmetry as a function of Cabibbo-Kobayashi-Maskawa angle $\alpha$. In the left panel, the solid line is for
$B^-\to  b^{0}_1\rho^-$ and the dashed line is for $B^-\to  a^{0}_1\rho^-$. In the right panel, the solid line is for $\bar B^0\to  b^{0}_1\rho^0$ and the dashed line
is for $\bar B^0\to  b^{0}_1\omega$. The vertical band shows the range of $\alpha$: $91.0\pm3.9$.}\label{fig4}
\end{center}
\end{figure}

Now we turn to the evaluations of the CP-violating asymmetries in PQCD approach. The CP asymmetries of $B^0/\bar B^0\to a^{\pm}_1(b^{\pm}_1)\rho^\mp$ are very
complicated and left for future study. Here we only research the decays $B^-\to  a^{0}_1(b^{0}_1)\rho^-$ and $\bar B^0\to  b^{0}_1\rho^0(\omega)$, where the
transverse polarization fractions are very small and range from $4.7$ to $7.5\%$.
Using Eq.(\ref{am}) and Eq.(\ref{amcon}), one can get the expression for the direct CP-violating asymmetry:
\be
\acp^{dir}&=&\frac{ |\overline{\cal M}|^2-|{\cal M}|^2 }{
 |{\cal M}|^2+|\overline{\cal M}|^2}=\frac{2z_L\sin\alpha\sin\delta_L}
{(1+2z_L\cos\alpha\cos\delta_L+z_L^2) }\;.
\en
Here for our considered four decays, the contributions from the transverse polarizations are very small, so we neglected them in our calculations.
Using the input parameters and the wave functions as specified in this section and Sec.\ref{proper}, one can find the PQCD predictions (in units of $10^{-2}$) for the direct CP-violating asymmetries of the
considered decays:
\be
\acp^{dir}(B^-\to  a^{0}_1\rho^-)&=&11.8^{+1.6+0.0}_{-1.4-0.0},\\
\acp^{dir}(B^-\to  b^{0}_1\rho^-)&=&-3.7^{+0.4+1.2}_{-0.3-1.2},\\
\acp^{dir}(\bar B^0\to  b^{0}_1\rho^0)&=&23.8^{+4.3+1.9}_{-4.2-1.9},\\
\acp^{dir}(\bar B^0\to  b^{0}_1\omega)&=&80.3^{+3.8+3.2}_{-4.8-3.2},
\en
where the errors are induced by the uncertainties of $B$ meson shape parameter $\omega_b=0.4\pm0.04$ and the threshold resummation parameter $c$, varying from $0.3$
to $0.4$.


\section{Conclusion}\label{summary}

In this paper, by using the decay constants and the light-cone distribution amplitudes
derived from QCD sum-rule method, we research  $B\to a_1(1260)\rho(\omega,\phi), b_1(1235)\rho(\omega,\phi)$
decays in PQCD factorization approach and find that
\begin{itemize}
\item
Except the decays $\bar B^0\to  a^{0}_1\rho^0(\omega)$, other tree-dominated
$B\to a_1\rho(\omega)$ decays have larger branching ratios, at the order of $10^{-5}$.
Except the decays $\bar B\to b^+_1\rho^-$ and $B^-\to b^0_1\rho^-$, other
$B\to b_1\rho(\omega)$ decays have smaller branching ratios, at the order of $10^{-6}$.
The decays $B\to a_1(b_1) \phi$ are highly suppressed and have very small
branching ratios, at the order of $10^{-9}$.
\item
For the decays $\bar B^0 \to a_1^0\rho^0$ and $B^-\to b_1^-\rho^0$, their two transverse polarizations are larger than
their longitudinal polarizations, which are about $43.3\%$ and $44.9\%$, respectively.
The two transverse polarization fractions have near values in the decays $B\to a_1\rho(\omega)$,
while have large differences in some of $B\to b_1\rho(\omega)$ decays.
\item
For the decays $B^-\to  a^{0}_1\rho^-, b^{0}_1\rho^-$ and
$\bar B^0\to  b^{0}_1\rho^0, b^{0}_1\omega$, where the transverse polarization fractions range from $4.7$ to $7.5\%$, we calculate their direct
CP-violating asymmetries with neglecting the transverse polarizations and find that those for two charged decays have smaller values, which are about $11.8\%$ and $-3.7\%$,
respectively.
\end{itemize}

\section*{Acknowledgment}
This work is partly supported by the National Natural Science Foundation of China under Grant No. 11147004, and by Foundation of Henan University of
Technology under Grant No. 2009BS038. The author would like to thank
Cai-Dian L\"u and Wei Wang for helpful discussions.

\end{document}